\documentclass{ws-ijmpa}

\usepackage[super]{cite}
\usepackage{xcolor}
\usepackage[verbose,hypertexnames=false]{hyperref}
\hypersetup{colorlinks=false,allbordercolors=blue,pdfborderstyle={/S/U/W 1}}

\begin{document}

\markboth{Yan Huang}{3D Lévy Femtoscopy via FFT in EPOS4, and Source-Dynamics Comparison with UrQMD}

%
\catchline{}{}{}{}{}
%

\title{Femtoscopic parameters of pion emission sources in EPOS4 Au+Au collisions from 7.7 to 200 GeV}

\author{Yan Huang, Máté Csanád}
\address{Department of Atomic Physics, ELTE Eötvös Loránd University\\
Pázmány Péter sétány 1/A, Budapest,H-1117, Hungry\\
h-y12@tsinghua.org.cn, csanad@elte.hu}

\maketitle

\begin{history}
\received{Day Month Year}
\revised{Day Month Year}
\accepted{Day Month Year}
\published{Day Month Year}
\end{history}

\begin{abstract}
Three-dimensional femtoscopy provides access to the space-time structure of the particle-emitting source in relativistic heavy-ion collisions. In particular, the collision-energy dependence of the Lévy source parameters may provide information on the evolution of strongly interacting matter and possible signatures associated with the QCD phase diagram. Recently, we studied the excitation function of three-dimensional Lévy source parameters of pion pairs in Au+Au collisions using the EPOS4 model, where the pair-distance distribution $D({\rho})$ was analyzed through its one-dimensional projections. In the present work, we extend this analysis by treating the full three-dimensional source distribution and correlation function directly. We construct D($\rho_{\rm out},\rho_{\rm side},\rho_{\rm long}$) as a three-dimensional histogram and transform it to the corresponding momentum-space correlation function $C(q_{\rm out},q_{\rm side},q_{\rm long})$ using a Fast Fourier Transform. The resulting three-dimensional correlation function is then fitted directly with the analytic Lévy formula. This procedure avoids possible biases associated with fitting only one-dimensional projections of a non-elliptical three-dimensional source. We apply the method to EPOS4 simulations over the RHIC Beam Energy Scan range and compare its resulting energy dependence with that obtained from the $D(\boldsymbol{\rho})$-based method. We then extend the study by comparing the energy dependence of the extracted source parameters between EPOS4 and recent UrQMD calculations. Particular attention is paid to the out-side observables $(R_{\rm diff}^{2}=R_{\rm out}^{2}-R_{\rm side}^{2})$ and $(R_{\rm out}/R_{\rm side})$, whose different energy dependences in EPOS4 and UrQMD provide insight into the role of the emission duration and the dynamical evolution of the source. This study establishes a more direct three-dimensional framework for future comparisons with experimental femtoscopic measurements and for investigations of possible non-monotonic structures related to the QCD phase diagram.
\end{abstract}

\keywords{Lévy source; Correlation function; HBT femtoscopy.}

\ccode{PACS numbers: 03.65.$-$w, 04.62.+v}

\section{Introduction}	
Relativistic heavy-ion collisions provide a unique opportunity to investigate strongly interacting matter under extreme conditions of temperature and baryon density. The space-time structure of the particle-emitting system can be studied with femtoscopy, which exploits the quantum-statistical correlations of identical particles at small relative momenta. In particular, two-pion femtoscopy provides information on the characteristic spatial and temporal scales of the source at kinetic freeze-out.

While Gaussian source parameterizations have traditionally been used, experimental measurements at different collision energies have demonstrated that L\'evy-stable distributions can provide a more flexible description of the non-Gaussian source structure~\cite{Porfy2024SPS,Kincses2024RHIC,Tumasyan2024LHC}. The L\'evy index $\alpha$ characterizes the shape of the source, with $\alpha=2$ corresponding to a Gaussian distribution and smaller values indicating increasingly non-Gaussian, long-tailed behavior. Such deviations from a Gaussian shape may originate from several physical mechanisms, including resonance decays, anomalous diffusion, and critical phenomena.

In particular, the L\'evy index has been proposed as a possible observable sensitive to the vicinity of the QCD critical point~\cite{Csorgo2006QCD}. Long-range correlations associated with critical phenomena may modify the spatial structure of the particle-emitting source and consequently affect the energy dependence of the L\'evy parameters. Therefore, systematic measurements of the collision-energy dependence of $\alpha$, the femtoscopic radii, and related observables are of interest for studies of the QCD phase diagram.

Three-dimensional femtoscopy is especially valuable because it separates the source structure into the Bertsch--Pratt out, side, and long directions. The corresponding radii ($R_{out}, R_{side}, R_{long}$) contain complementary information about the transverse geometry, longitudinal expansion, and emission dynamics~\cite{Kincses2026}. In particular, the difference between the out and side directions is sensitive to the duration of particle emission and has been discussed as a possible probe of a first-order phase transition~\cite{Csanád_2025}.

Recently, we investigated the excitation function of three-dimensional L\'evy source parameters in EPOS4 for Au+Au collisions at
$\sqrt{s_{NN}}=7.7-200~{\rm GeV}$~\cite{Huang2026}.
The study showed that $R_{\rm side}$ and $R_{\rm long}$ increase with collision energy, while $R_{\rm out}$ has a much weaker energy dependence. The L\'evy index exhibits only a mild energy dependence, while the correlation strength $\lambda$ decreases with increasing collision energy. The derived quantities
$R_{\rm diff}^{2}=R_{\rm out}^{2}-R_{\rm side}^{2}$ 
and $R_{\rm out}/R_{\rm side}$ showed characteristic but different energy dependences. These results provide a non-critical baseline for future experimental measurements.

However, the previous analysis extracted the three-dimensional L\'evy parameters by fitting the one-dimensional projections of the three-dimensional pair-distance distribution. This approach is useful and computationally efficient, but it implicitly relies on the assumption that the information contained in the projections is sufficient to characterize the underlying three-dimensional source.
A direct inspection of the multidimensional source distributions indicates that this assumption is not always adequate. In particular, the three-dimensional $D({\rho})$ distributions can exhibit visibly non-elliptical structures. In such a situation, different three-dimensional source shapes can lead to similar one-dimensional projections, potentially introducing a bias when the source parameters are extracted only from the projections.

The aim of the present work is therefore twofold. First, we optimize the
$C(\mathbf{q})$ method by developing a direct three-dimensional analysis in which the measured pair-distance distribution $D(\boldsymbol{\rho})$ is transformed into the momentum-space correlation function $C(\mathbf{q})$ using a Fast Fourier Transform (FFT), and the full three-dimensional correlation function is fitted directly. Second, within the $D(\boldsymbol{\rho})$ framework, we revisit the energy dependence of the L\'evy source parameters and investigate the difference
between hydrodynamic and transport descriptions through a comparison of EPOS4 and UrQMD.

This approach provides a direct connection between the emission source and the experimentally accessible correlation function and may serve as a more robust framework for future three-dimensional femtoscopic studies.

\section{Three-dimensional source and correlation function}

\subsection{Pair-distance distribution}
In a transport or event-generator calculation, the relative space-time coordinates of two emitted particles can be used to construct the pair-distance distribution~\cite{Kincses:2024lnv}
\begin{align}
\rho_{\mathrm{out}}^{\mathrm{LCMS}} &= r_x \cos\varphi + r_y \sin\varphi 
- \frac{k_T}{K_0^2 - K_z^2} \left( K_0 t - K_z r_z \right), \\
\rho_{\mathrm{side}}^{\mathrm{LCMS}} &= - r_x \sin\varphi + r_y \cos\varphi, \\
\rho_{\mathrm{long}}^{\mathrm{LCMS}} &= \frac{K_0 r_z - K_z t}{\sqrt{K_0^2 - K_z^2}}.
\end{align}
where the coordinates are defined in the longitudinally co-moving system (LCMS) using the Bertsch--Pratt decomposition.
The out direction is defined by the transverse momentum of the pair, the side direction is perpendicular to the out direction in the transverse plane, and the long direction is parallel to the beam axis.

For a L\'evy-stable source, the three-dimensional pair-distance distribution can be written as~\cite{Nolan2013LevyEq}:
\begin{equation}
D(\boldsymbol{\rho})
=
\frac{1}{(2\pi)^3}
\int d^3\boldsymbol{\zeta}\,
e^{i\boldsymbol{\zeta}\cdot\boldsymbol{\rho}}
\exp\left[
-\left|
\boldsymbol{\zeta}^{T}R^2\boldsymbol{\zeta}
\right|^{\alpha/2}
\right],
\end{equation}
where
$R^2 ={\rm diag}\left(R_{\rm out}^2,R_{\rm side}^2,R_{\rm long}^2\right)$.
The L\'evy index $\alpha$ characterizes the source shape, while the three scale parameters describe the source size in the three Bertsch--Pratt directions. In the previous EPOS4 analysis, one-dimensional projections of the three-dimensional source were fitted simultaneously to extract these parameters. In the present work, instead of reducing the source to its projections, we retain the complete three-dimensional information.

\subsection{Construction of the correlation function}

The experimentally measured observable is the momentum-space two-particle correlation function,
\begin{equation}
C(\mathbf{q})
=
\int d^3\boldsymbol{\rho}\,
D(\boldsymbol{\rho})
\left|
\psi_{\mathbf{q}}(\boldsymbol{\rho})
\right|^2.
\end{equation}

For the present study, we focus on the quantum-statistical component and use the Fourier relationship between the source and correlation functions. The three-dimensional $D(\boldsymbol{\rho})$ distribution is stored directly as a TH3F histogram.
For each collision energy, transverse-momentum interval, and centrality class, the pair-distance distributions are accumulated over events. The centrality classes and the definition of transverse mass are the same as those used in the previous EPOS4 study~\cite{Huang2026}, with $m_T=\sqrt{m^2+k_T^2}$,
where $m$ is the mass of the investigated particle and $k_T$ is the transverse momentum. The distribution is normalized using the full histogram range, including the underflow and overflow bins, such that the integral over the complete source distribution is properly normalized. 

The three-dimensional histogram is then transformed into momentum space using a Fast Fourier Transform~\cite{fftw}. This produces a numerical representation of $C(q_{\rm out},q_{\rm side},q_{\rm long})$
on a uniform three-dimensional momentum grid.
This procedure retains the complete multidimensional structure of the source and avoids reducing the distribution to one-dimensional projections before parameter extraction.

The transform is performed on a cubic grid with $N$ bins along each spatial direction. The same odd number of bins is used in all three directions so that the central bin corresponds exactly to $\boldsymbol{\rho}=0$. Before the transform, the histogram is normalized by its full integral, including the underflow and overflow bins,
\begin{equation}
\tilde{D}(\boldsymbol{\rho}_i)
=
\frac{D(\boldsymbol{\rho}_i)}
{\sum_{i,j,k=0}^{N+1} D(\rho_{x,i},\rho_{y,j},\rho_{z,k})},
\end{equation}
so that the source is normalized to unity over the complete histogram range.
The normalized source is then discretized on this grid and Fourier transformed,
\begin{equation}
F(\mathbf{q}_m)
=
\sum_{n=0}^{N-1}
\tilde{D}(\boldsymbol{\rho}_n)\,
e^{\,+\,2\pi i\, \mathbf{m}\cdot\mathbf{n}/N},
\end{equation}
where $\mathbf{n}=(n_x,n_y,n_z)$ labels the spatial grid points and $\mathbf{m}=(m_x,m_y,m_z)$ labels the momentum grid points. The integer indices are mapped to physical momenta through
\begin{equation}
m_x =
\begin{cases}
i, & i \le (N-1)/2,\\
i-N, & i > (N-1)/2,
\end{cases}
\end{equation}
and analogously for $m_y$ and $m_z$. The momentum spacings are determined by the spatial bin widths,
\begin{equation}
\Delta q_{\rm out}
=
\frac{2\pi}{N\,\Delta\rho_{\rm out}}\,\hbar c,
\qquad
\Delta q_{\rm side}
=
\frac{2\pi}{N\,\Delta\rho_{\rm side}}\,\hbar c,
\qquad
\Delta q_{\rm long}
=
\frac{2\pi}{N\,\Delta\rho_{\rm long}}\,\hbar c,
\end{equation}
with $\hbar c = 0.197327\ \mathrm{GeV\,fm}$, so that the momentum grid is centered around zero and spans both positive and negative $q$.

\subsection{Three-dimensional L\'evy fit of $C(\mathbf{q})$}

The resulting three-dimensional correlation function is fitted with the analytic form
\begin{equation}
C(\mathbf{q})
=
1+\lambda
\exp\left[
-\left|
\mathbf{q}^{T}R^{2}\mathbf{q}
\right|^{\alpha/2}
\right],
\end{equation}
The five fit parameters are therefore
$\alpha, \lambda, R_{\rm out}, R_{\rm side}, R_{\rm long}$.
Because the FFT-generated correlation function has approximately uniform uncertainties, the absolute statistical uncertainties obtained directly from the fit do not necessarily correspond to the expected uncertainty scale associated with the fit quality. We therefore rescale the parameter uncertainties according to the fitted $\chi^2/{\rm NDF}$, such that the final uncertainties correspond to a $\chi^2$ scaled to the NDF.

Different fitting ranges are tested to evaluate the stability of the results. In addition to a cubic range, defined by independent limits on $q_{\rm out}$, $q_{\rm side}$, and $q_{\rm long}$, a spherical range, defined by $q_{\rm min}^2<q_{\rm out}^2+q_{\rm side}^2+q_{\rm long}^2<q_{\rm max}^2$, is considered. Different binning schemes are also tested. The results obtained with these different binning and fitting-range selections exhibit only minor numerical variations and are consistent within the quoted uncertainties.

\begin{figure}[t]
\centerline{\includegraphics[width=\columnwidth]{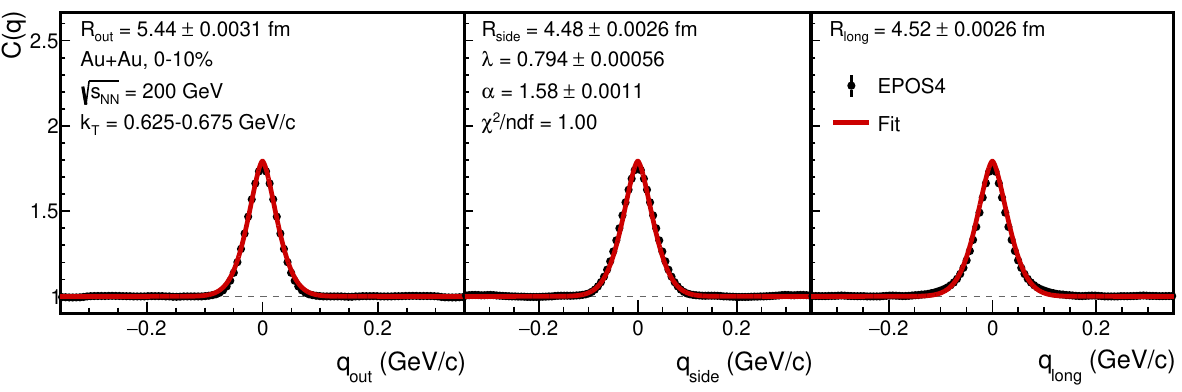}}
\caption{Projections of the 3D $C(\mathbf{q})$ fit in the out, side, and long directions for $0-10\%$ Au+Au collisions at $\sqrt{s_{NN}}=200$ GeV in the highest analyzed $k_T$ bin, $0.625<k_T<0.675$ GeV/$c$.
}
\label{fit}
\alttext{fit}
\end{figure}

Some statistical fluctuations are visible in the higher-$k_T$ bins at relatively large $q$ values, up to about $0.6$--$0.8$~GeV/$c$, due to the limited statistics. 
However, within the fitting range of approximately $q<0.3$--$0.4$~GeV/$c$, the $C(\mathbf{q})$ projections remain smooth, as illustrated in Fig.~\ref{fit}. For most of the other lower-$k_T$ bins, the $C(\mathbf{q})$ projections are smooth over the entire shown $q$ range.

\section{Results and discussions}

The analysis is performed for Au+Au collisions generated with EPOS4 in the RHIC Beam Energy Scan energy range. The event and particle selections follow the setup used in our previous study, including identical pion pairs and the STAR-inspired kinematic acceptance. Unlike in our previous analysis, where events were merged, all generated events are analyzed individually here to maximize the statistical precision.

By transforming the complete $D(\boldsymbol{\rho})$ distribution into $C(\mathbf{q})$, the analysis retains the full multidimensional structure and allows the L\'evy parameterization to be tested directly in momentum space.
The comparison between the projection-based and direct three-dimensional analyses will therefore provide an important consistency test of the previously extracted excitation functions.

In this section, the parameters extracted from correlation functions $C(\mathbf{q})$ are compared with that from pair distance $D(\boldsymbol{\rho})$. In particular, we investigate whether the values and energy dependence of parameters
remain stable when the complete three-dimensional correlation function is fitted.

\subsection{Source shape}

The L\'evy index showed only a mild energy dependence in EPOS4, with no clear non-monotonic structure over the investigated energy range. This result is important because EPOS4, with its standard crossover equation of state, does not contain an explicit critical point. Its energy dependence can therefore be regarded as a non-critical baseline against which future calculations with different equations of state and future experimental measurements can be compared.

\begin{figure}[t]
\centerline{\includegraphics[width=\columnwidth]{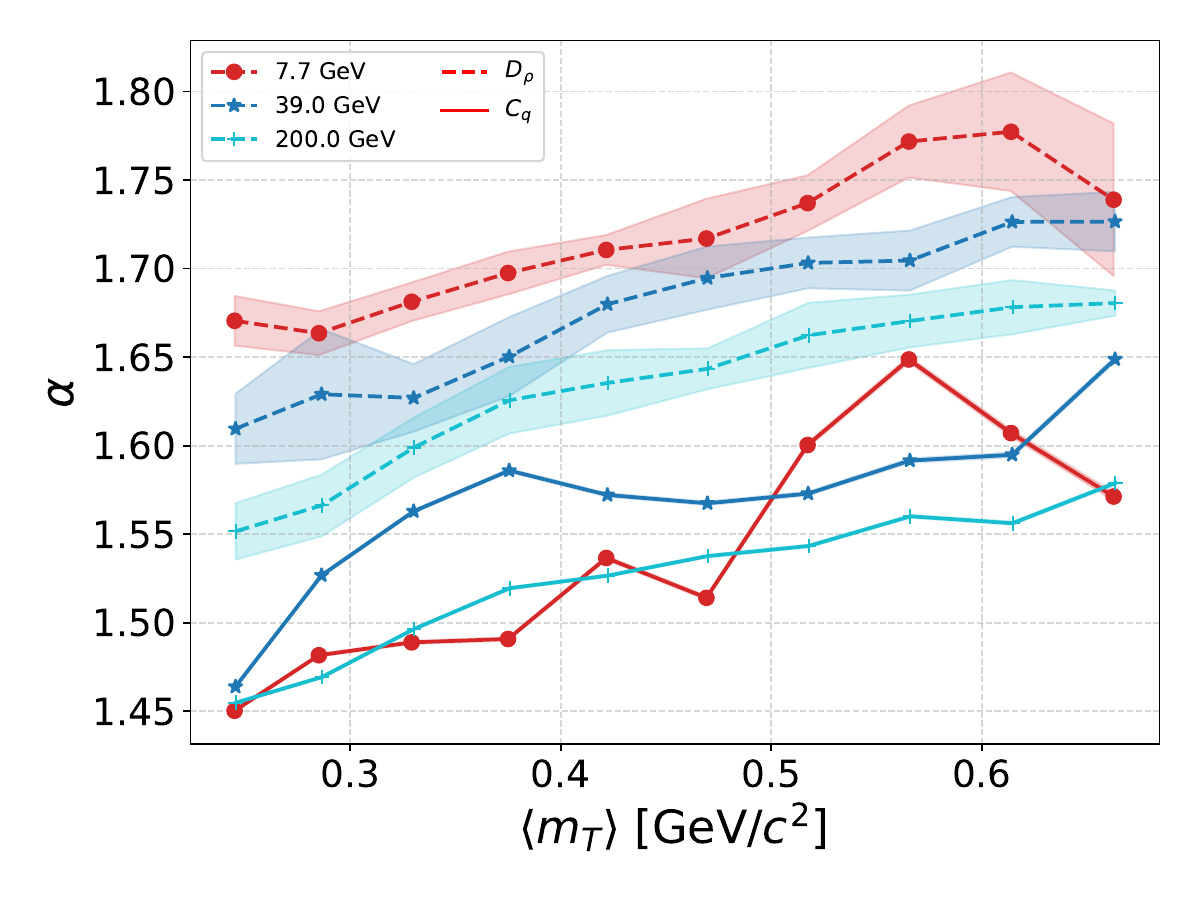}}
\caption{Lévy index $\alpha$ obtained from the $C(\mathbf{q})$ FFT analysis of EPOS4 simulations for $0$--$10\%$ Au+Au collisions as a function of transverse mass $m_T$ at energies 7.7, 39 and 200~GeV. The solid lines represent the $C(\mathbf{q})$ FFT results, while the dashed lines represent the $D\boldsymbol{\rho})$-based results from EPOS4. The bands show the systematic uncertainties. }
\label{fig:f1}
\alttext{alpha_mT}
\end{figure}

Fig.~\ref{fig:f1} shows the $m_T$ dependence of $\alpha$ at three representative collision energies, 7.7, 39.0, and 200.0 GeV. The $C(\mathbf{q})$ fits yield systematically smaller values of $\alpha$ than the $D(\boldsymbol{\rho})$-based method, with differences of less than 15\% at 7.7 and 9.2~GeV and less than 10\% at all other energies.

\begin{figure}[t]
\centerline{\includegraphics[width=\columnwidth]{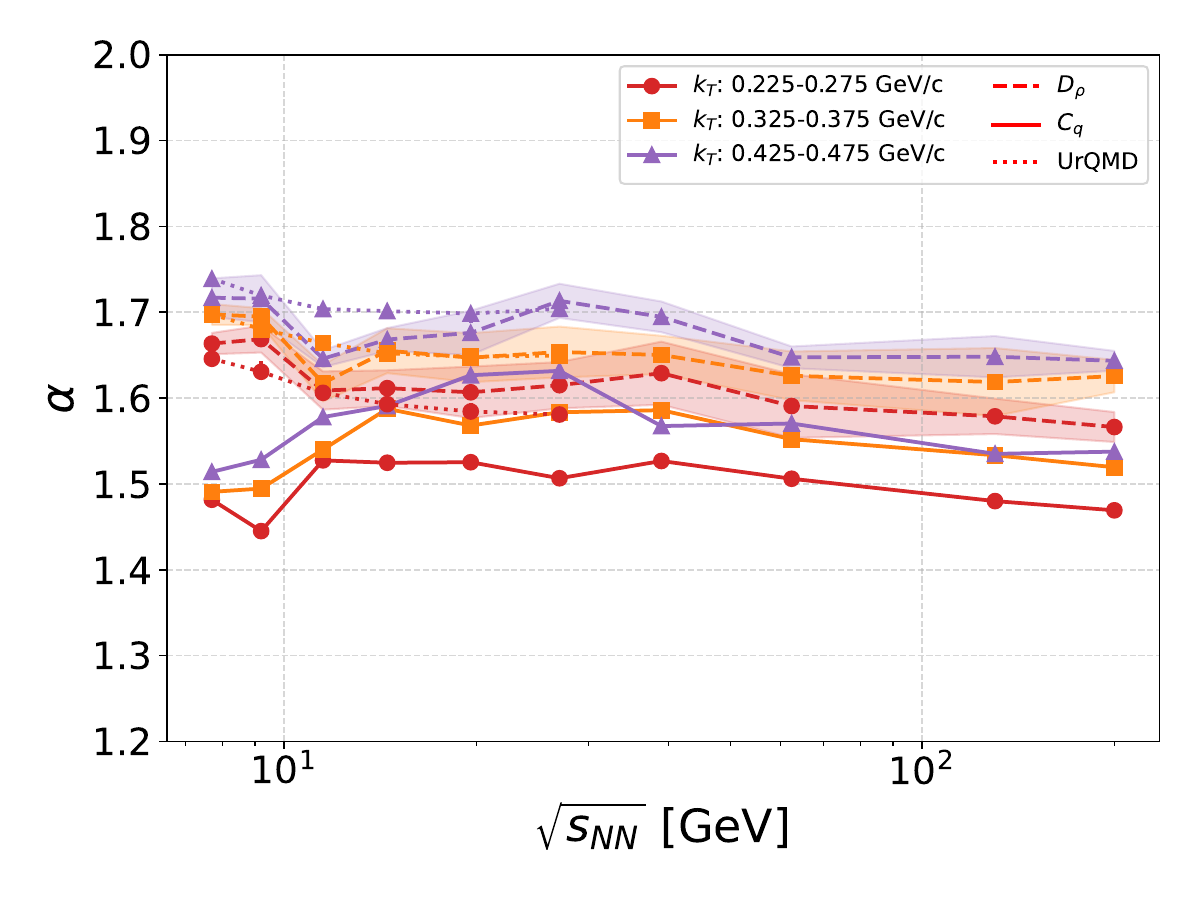}}
\caption{Lévy index $\alpha$ obtained from the $C(\mathbf{q})$ FFT analysis of EPOS4 simulations for $0$--$10\%$ Au+Au collisions as a function of collision energy $\sqrt{s_{NN}}$ in three representative $k_T$ bins. The bands show the systematic uncertainties. The solid, dashed, and dotted lines represent the $C(\mathbf{q})$ FFT results, the $D(\boldsymbol{\rho})$-based results from EPOS4, and the $D(\boldsymbol{\rho})$-based results from UrQMD~\cite{Molnar2026}, respectively.}
\label{fig:f2}
\alttext{alpha_energy}
\end{figure}

Fig.~\ref{fig:f2} shows the energy dependence of $\alpha$ in three $k_T$ bins, $0.375$--$0.425$, $0.425$--$0.475$, and $0.475$--$0.525$ GeV/$c$, corresponding to mean $m_T$ values of 0.422, 0.469, and 0.517 GeV/$c^2$, respectively. For completeness, the Urqmd results  obtained with the $D(\boldsymbol{\rho})$-based method ~\cite{Molnar2026} are also shown as dotted lines in the figures.These results are included for comparison but are not discussed further in this section; a detailed comparison of the EPOS4 and UrQMD results is presented in Sec.~\ref{sec:EPOS4_UrQMD}. 
As shown in the figures, the $C(\mathbf{q})$ fits reproduce the similar $m_T$ and energy dependence of $\alpha$, while yielding systematically lower values than those obtained with the $D(\boldsymbol{\rho})$-based method.

\subsection{Source size}
In the previous EPOS4 study, the radii $R_{\rm side}$ and $R_{\rm long}$ were found to increase with collision energy, while $R_{\rm out}$ exhibited a comparatively weak energy dependence. The increase of $R_{\rm long}$ is consistent with a longer lifetime and stronger longitudinal development of the system at higher energies, while the increase of $R_{\rm side}$ reflects the evolution of the transverse source size~\cite{Huang2026}.

\begin{figure}[t]
\centerline{\includegraphics[width=\columnwidth]{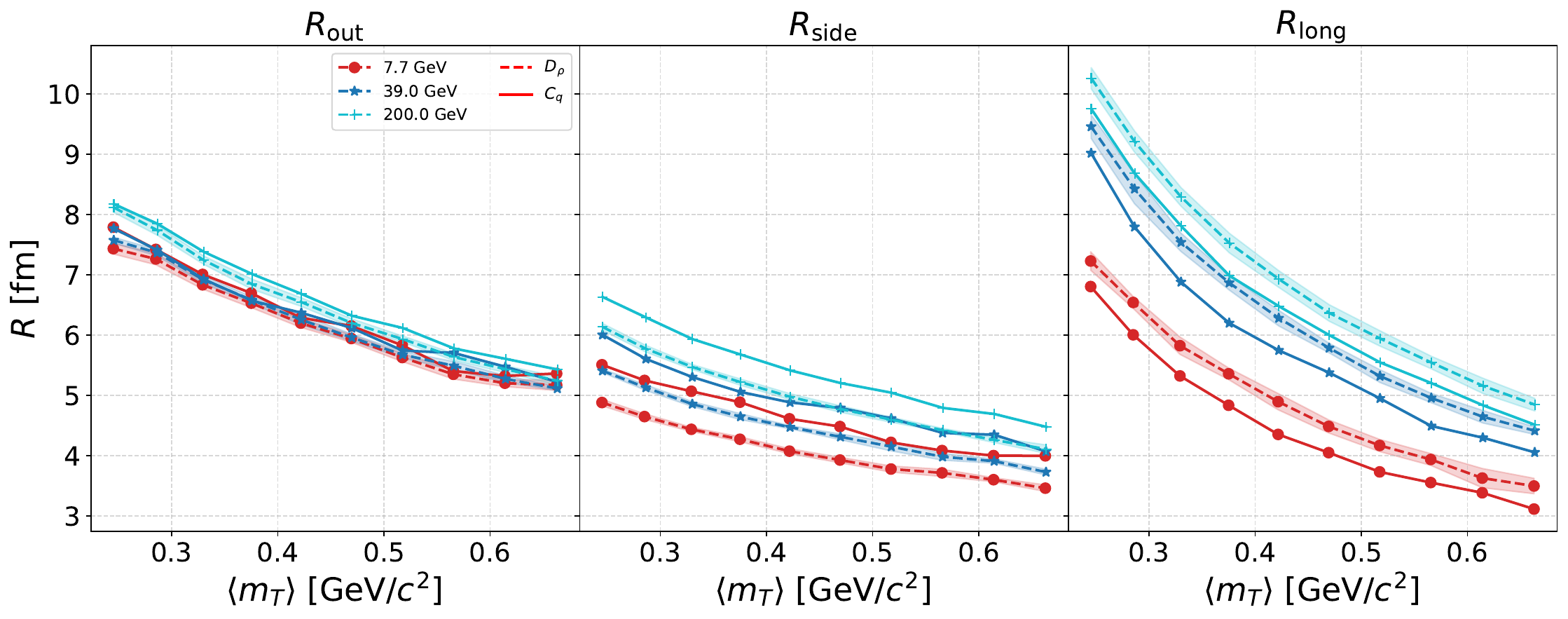}}
\caption{Lévy radii obtained from the $C(\mathbf{q})$ FFT analysis of EPOS4 simulations for $0-10\%$ Au+Au collisions as a function of transverse mass $m_T$ at energies 7.7, 39 and 200GeV. The solid lines represent the $C(\mathbf{q})$ FFT results, while the dashed lines represent the $D(\boldsymbol{\rho})$-based results from EPOS4. The bands show the systematic uncertainties. }
\label{Radii_mT}
\alttext{Radii_mT}
\end{figure}

The $R_{\rm out}$ values obtained from the $C(\mathbf{q})$ fits are in close agreement with those from the $D(\boldsymbol{\rho})$-based method, with differences of less than 5\%. In comparison, $R_{\rm side}$ is systematically larger by about 5--20\%, whereas $R_{\rm long}$ is systematically smaller by about 5--15\%, as shown in Fig.~\ref{Radii_mT}.

\begin{figure}[t]
\centerline{\includegraphics[width=\columnwidth]{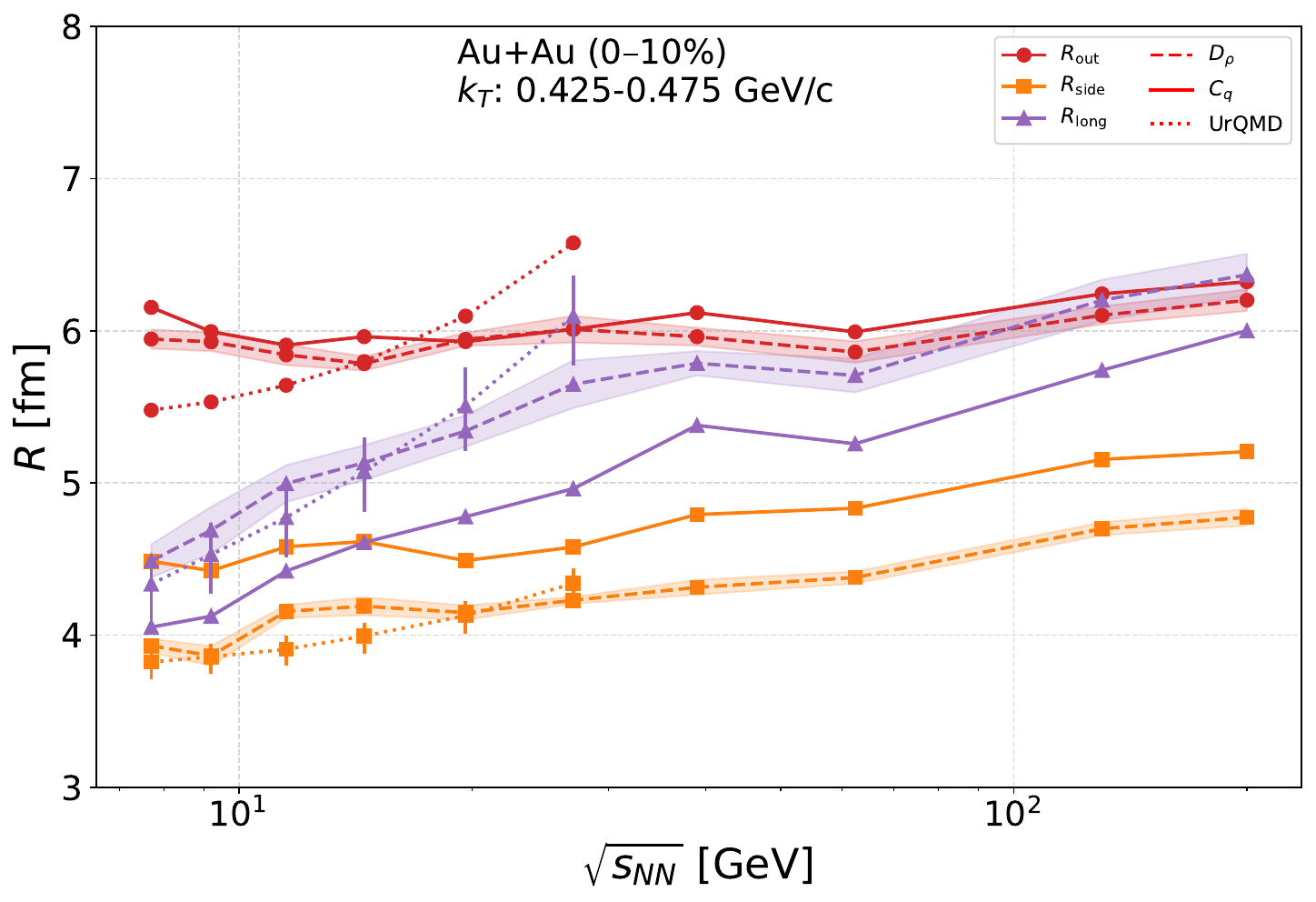}}
\caption{Lévy radii obtained from the $C(\mathbf{q})$ FFT analysis of EPOS4 simulations for $0-10\%$ Au+Au collisions as a function of collision energy $\sqrt{s_{NN}}$ in three representative $k_T$ bins. The solid, dashed, and dotted lines represent the $C(\mathbf{q})$ FFT results, the $D(\boldsymbol{\rho})$-based results from EPOS4, and the $D(\boldsymbol{\rho})$-based results from UrQMD~\cite{Molnar2026}, respectively.}
\label{Radii_Ecm}
\alttext{Radii_mT}
\end{figure}

\begin{figure}[t]
\centerline{\includegraphics[width=\columnwidth]{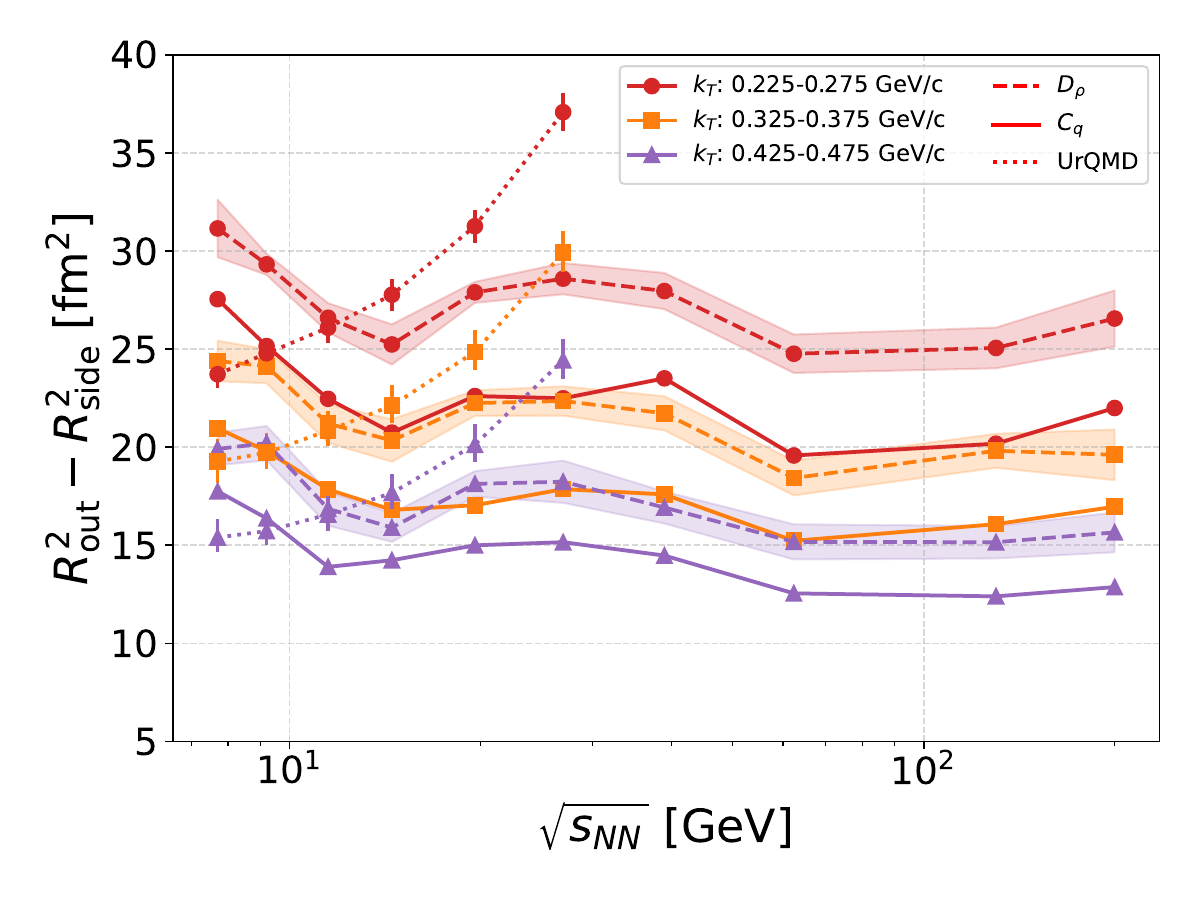}}
\caption{The radius difference $R_\mathrm{diff}^2$ from the $C(\mathbf{q})$ FFT analysis of EPOS4 simulations for $0$--$10\%$ Au+Au collisions as a function of collision energy $\sqrt{s_{NN}}$ in three representative $k_T$ bins. The solid, dashed, and dotted lines represent the $C(\mathbf{q})$ FFT results, the $D(\boldsymbol{\rho})$-based results from EPOS4, and the $D(\boldsymbol{\rho})$-based results from UrQMD~\cite{Molnar2026}, respectively. The bands show the systematic uncertainties.}
\label{fig:Rdiff_Ecm}
\alttext{Rdiff_Ecm}
\end{figure}

\begin{figure}[t]
\centerline{\includegraphics[width=\columnwidth]{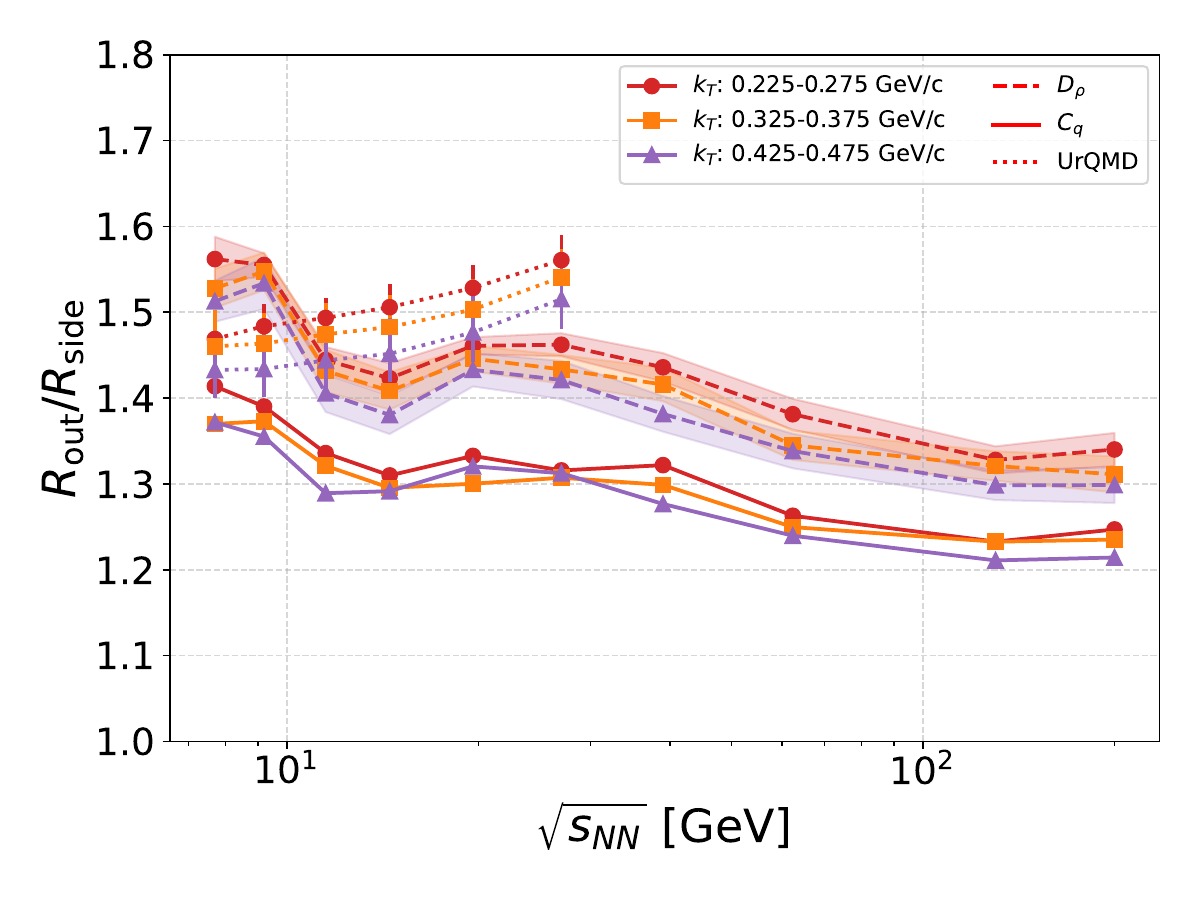}}
\caption{The radius ratio $R_\mathrm{out}/R_\mathrm{side}$ from the $C(\mathbf{q})$ FFT analysis of EPOS4 simulations for $0$--$10\%$ Au+Au collisions as a function of collision energy $\sqrt{s_{NN}}$ in three representative $k_T$ bins. The solid, dashed, and dotted lines represent the $C(\mathbf{q})$ FFT results, the $D(\boldsymbol{\rho})$-based results from EPOS4, and the $D(\boldsymbol{\rho})$-based results from UrQMD~\cite{Molnar2026}, respectively. The bands show the systematic uncertainties.}
\label{fig:Rratio_Ecm}
\alttext{Rratio_energy}
\end{figure}

The collision-energy dependence of all three Lévy radii is consistent with that obtained using the $D(\boldsymbol{\rho})$-based method.The same overall energy-dependent trends are reproduced for $R_{\rm out}$, $R_{\rm side}$, and $R_{\rm long}$, as shown in Fig.~\ref{Radii_Ecm}. The corresponding observables $R_{\rm diff}^{2}$ and $R_{\rm ratio}$ also exhibit energy dependences that are in good agreement with the $D(\boldsymbol{\rho})$-based results, as shown in Fig.~\ref{fig:Rdiff_Ecm} and Fig.~\ref{fig:Rratio_Ecm}, respectively.

\subsection{Correlation strength}

\begin{figure}[t]
\centerline{\includegraphics[width=\columnwidth]{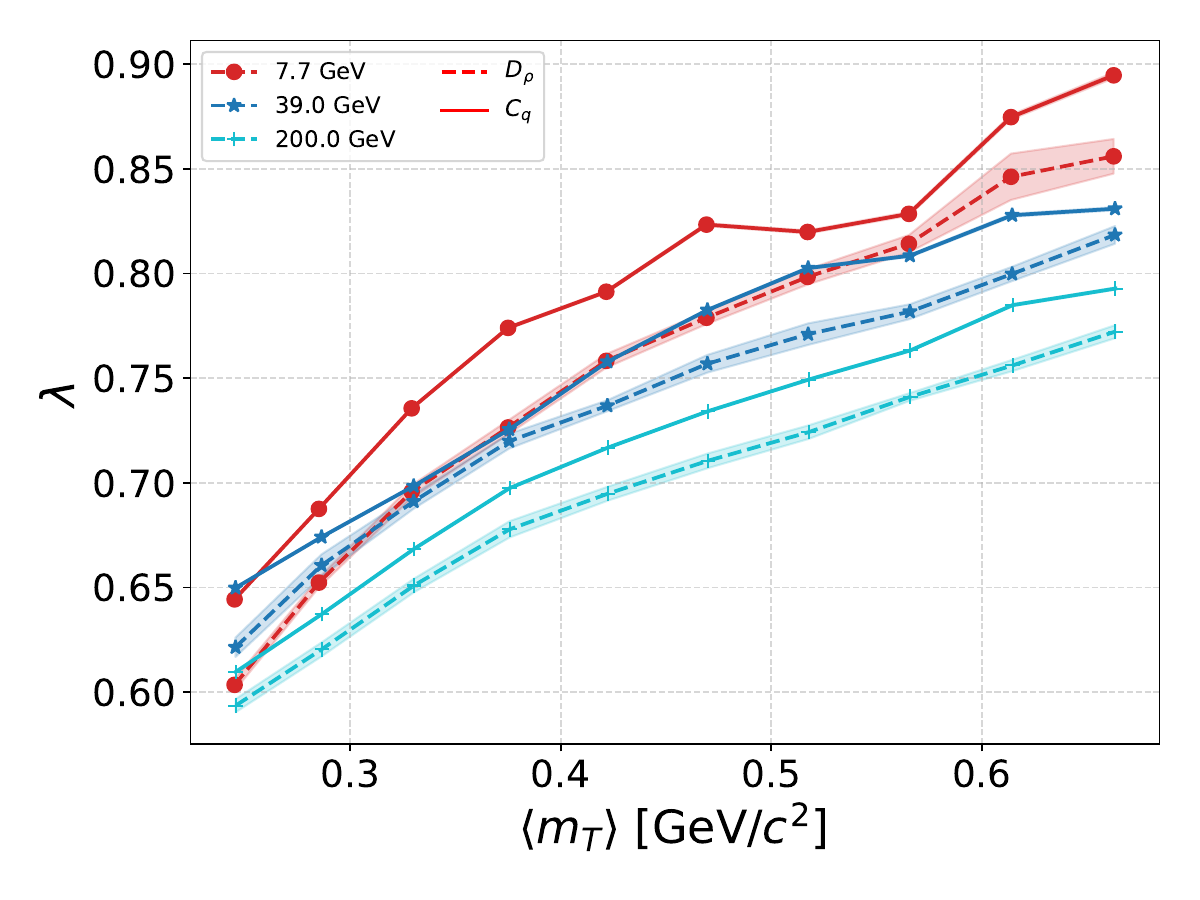}}
\caption{Correlation strength parameters $\lambda$ from the $C(\mathbf{q})$ FFT analysis of EPOS4 simulations for $0$--$10\%$ Au+Au collisions as a function of transverse mass $m_T$ at energies 7.7, 39 and 200~GeV. The solid lines represent the $C(\mathbf{q})$ FFT results, while the dashed lines represent the $D(\boldsymbol{\rho})$-based results from EPOS4. The bands show the systematic uncertainties.}
\label{fig:lambda_mT}
\alttext{lambda_mT}
\end{figure}

\begin{figure}[t]
\centerline{\includegraphics[width=\columnwidth]{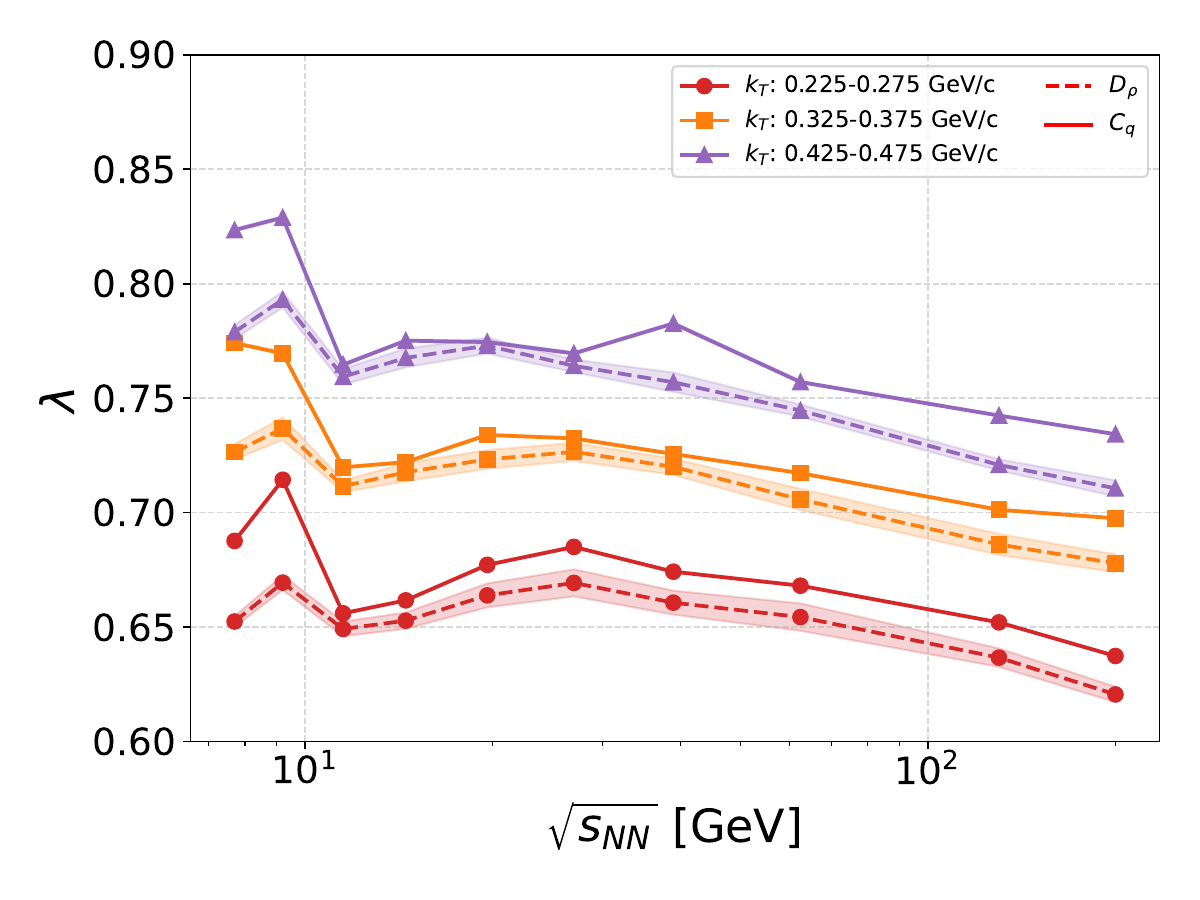}}
\caption{Correlation strength parameters $\lambda$ from the $C(\mathbf{q})$ FFT analysis of EPOS4 simulations for $0$--$10\%$ Au+Au collisions as a function of collision energy $\sqrt{s_{NN}}$ in three representative $k_T$ bins. The solid and dashed lines represent the $C(\mathbf{q})$ FFT results, and the $D(\boldsymbol{\rho})$-based results from EPOS4. The bands show the systematic uncertainties.}
\label{fig:lambda_Ecm}
\alttext{lambda_energy}
\end{figure}

The $\lambda$ values obtained from the $C(\mathbf{q})$ fits are generally systematically larger than those from the $D(\boldsymbol{\rho})$-based analysis, with most differences within 0--7\% and only a few points showing differences of less than -2\%. Their $m_T$ and collision-energy dependences are in good agreement, as shown in Figs.~\ref{fig:lambda_mT} and \ref{fig:lambda_Ecm}.

Overall, The Lévy index $\alpha$ shows a systematic difference between the $C(\mathbf{q})$- and $D(\boldsymbol{\rho})$-based analyses. While the difference is moderate over most of the collision-energy range, it becomes more pronounced at the two lowest energies, 7.7 and 9.2 GeV. In contrast, the extracted values of $\lambda$ and the three Lévy radii show no corresponding anomalous behavior, and their collision-energy dependences remain consistent with those obtained from the $D(\boldsymbol{\rho})$-based analysis. 

\section{Comparison between EPOS4 and UrQMD}
\label{sec:EPOS4_UrQMD}

To further investigate the collision-energy dependence and its sensitivity to the underlying collision dynamics, we compare our EPOS4 results with a recent UrQMD calculation of pion emission source parameters in Au+Au collisions~\cite{Molnar2026}. Both results are obtained using the $D(\boldsymbol{\rho})$-based analysis method, ensuring a direct comparison of the extracted source parameters. UrQMD provides a complementary description of the collision dynamics based on a hadronic transport approach, without an explicit hydrodynamic phase or equation of state. The corresponding analysis covers the collision-energy range from 3 to 27~GeV and extracts the same basic Lévy source parameters.

As shown in Fig~\ref{fig:f2},the $\alpha$ values obtained from UrQMD and EPOS4 are very close and show good agreement over the common collision-energy range.

As shown in Fig~\ref{Radii_Ecm}, the $R_{\rm out}$ values from UrQMD exhibit a clear increase with collision energy, whereas $R_{\rm out}$ from EPOS4 shows a much weaker energy dependence. Nevertheless, the average values are similar over the overlapping energy range of 7.7--27~GeV. The $R_{\rm long}$ values from UrQMD show a more pronounced increase with collision energy than those from EPOS4, while $R_{\rm side}$ exhibits a slight increase with energy in both models.

The derived observables $R_{\rm diff}^{2}=R_{\rm out}^{2}-R_{\rm side}^{2}$ and $R_{\rm out}/R_{\rm side}$ show different energy dependences in the two models. In particular, the EPOS4 results decrease with collision energy, whereas the UrQMD results increase. The two trends cross within the overlapping energy range, while their average values remain similar.

The different energy dependences of $R_{\rm diff}^{2}$ and $R_{\rm out}/R_{\rm side}$ can be largely traced to the different collision-energy dependence of $R_{\rm out}$ in EPOS4 and UrQMD. While $R_{\rm side}$ shows a similar, moderate increase with collision energy in both models, $R_{\rm out}$ exhibits a much stronger increase in UrQMD, whereas its energy dependence is relatively weak in EPOS4. Consequently, $R_{\rm diff}^{2}=R_{\rm out}^{2}-R_{\rm side}^{2}$ increases with collision energy in UrQMD but shows a decreasing trend in EPOS4. Similarly, the ratio $R_{\rm out}/R_{\rm side}$ increases in UrQMD and decreases in EPOS4.

\begin{figure}[t]
\centerline{\includegraphics[width=\columnwidth]{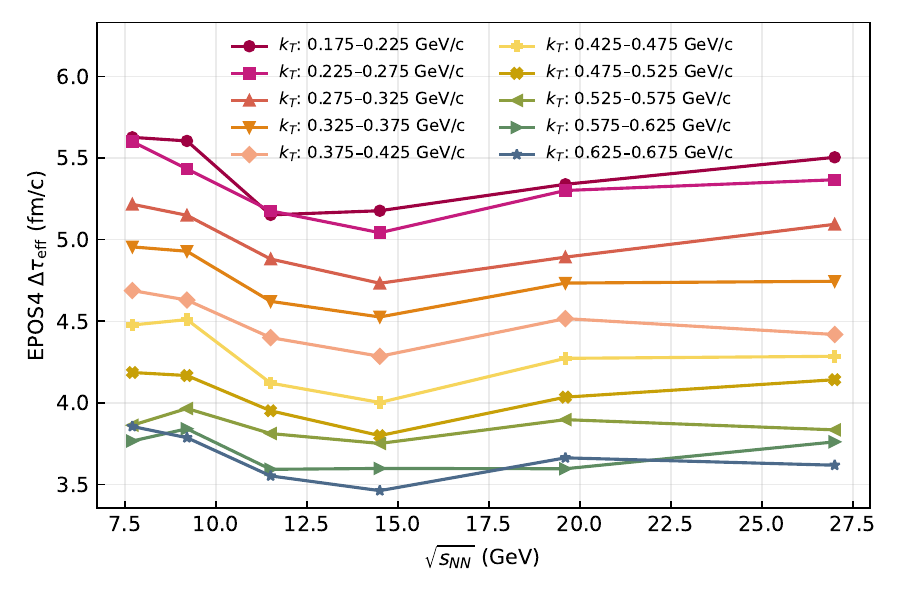}}
\caption{The effective emission-time scale $\Delta\tau_{\rm eff}$ obtained from the $D(\boldsymbol{\rho})$-based analysis of EPOS4 simulations for $0$--$10\%$ Au+Au collisions as a function of collision energy $\sqrt{s_{NN}}$ for all analyzed $k_T$ bins.}
\label{tau_epos4}
\alttext{tau_epos4}
\end{figure}

\begin{figure}[t]
\centerline{\includegraphics[width=\columnwidth]{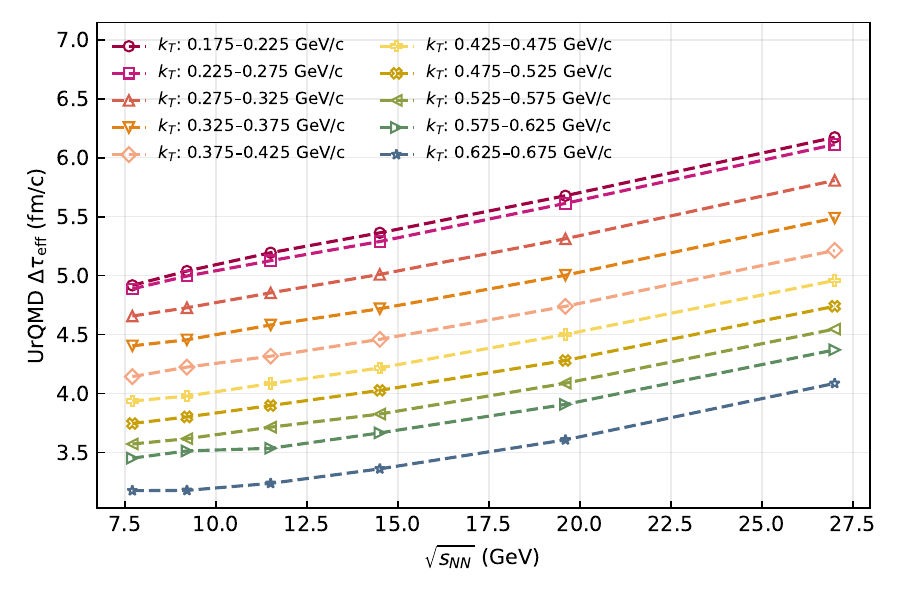}}
\caption{The effective emission-time scale $\Delta\tau_{\rm eff}$ obtained from the $D(\boldsymbol{\rho})$-based analysis of UrQMD simulations~\cite{Molnar2026} for $0$--$10\%$ Au+Au collisions as a function of collision energy $\sqrt{s_{NN}}$ for all analyzed $k_T$ bins.}
\label{tau_urqmd}
\alttext{tau_urqmd}
\end{figure}

The difference in $R_{\rm diff}^{2}$ can also be related to the emission-time scale. In the commonly used approximation
\begin{equation}
R_{\rm out}^{2}-R_{\rm side}^{2}\simeq \beta_T^2\Delta\tau^2,
\end{equation}
an effective emission duration can be estimated as
\begin{equation}
\Delta\tau_{\rm eff}
\simeq
\frac{\sqrt{R_{\rm out}^{2}-R_{\rm side}^{2}}}{\beta_T}.
\end{equation}
Using this relation, we find that the effective emission duration in EPOS4 shows only a weak dependence on collision energy, whereas the UrQMD results exhibit a pronounced, approximately linear increase with $\sqrt{s_{NN}}$. As shown in Fig~\ref{tau_epos4} and Fig~\ref{tau_urqmd}. This indicates that the two models predict different collision-energy dependences of the characteristic emission-time scale. At a fixed collision energy, the extracted effective emission-time scale decreases with increasing \(k_T\) in both models. This behavior is consistent with the \(k_T\)-dependent reduction of the homogeneity region: higher-\(k_T\) particles preferentially probe a more localized region of the evolving source. The observed \(k_T\) dependence therefore reflects the temporal extent of the homogeneity region rather than necessarily indicating a shorter global emission duration.

It should be noted that $R_{\rm out}^{2}-R_{\rm side}^{2}$ is not determined solely by the emission duration, as space--time correlations between the emission position and emission time can also contribute. Therefore, $\Delta\tau_{\rm eff}$ should be interpreted as an effective emission-time scale rather than a direct measurement of the emission duration. Nevertheless, the different energy dependences observed in $\Delta\tau_{\rm eff}$ provide an indication that the underlying space--time evolution of the pion-emitting source differs between EPOS4 and UrQMD.

The comparison with UrQMD therefore provides an important dynamical interpretation of the out-side observables. In the hydrodynamic description of EPOS4, the stronger collective evolution and the presence of a hydrodynamic phase modify the relation between the spatial source size and the emission duration. In contrast, in UrQMD the collision system evolves through hadronic rescattering, and the increasing duration of the cascade at higher multiplicity can lead to a stronger outward source component.This difference naturally leads to different excitation functions. 

Importantly, a non-monotonic structure in an out-side observable does not by itself imply critical-point dynamics. The excitation functions of femtoscopic observables can be affected by the underlying collision dynamics and space-time evolution of the emitting source, including collective expansion, emission duration, and space-time correlations. Although the two models give similar values of the Lévy index $\alpha$, they exhibit substantially different collision-energy dependences of $R_{\rm out}$ and, consequently, of $R_{\rm diff}^{2}$ and $R_{\rm out}/R_{\rm side}$.  These differences arise within the respective model descriptions of the collision evolution and demonstrate that the excitation functions of femtoscopic observables are sensitive to the underlying space-time dynamics. EPOS4 and UrQMD therefore provide complementary baseline calculations for studying the collision-energy dependence of the source parameters. Comparing experimental excitation functions with such baseline expectations is therefore an important step toward identifying structures that may be associated with a phase transition or, more specifically, with critical-point-related phenomena.

\section{Summary and outlook}

We have developed a direct three-dimensional approach to Lévy femtoscopy in which the complete pair-distance distribution
$D(\rho_{\rm out},\rho_{\rm side},\rho_{\rm long})$
is transformed into the momentum-space correlation function
$C(q_{\rm out},q_{\rm side},q_{\rm long})$ using a Fast Fourier Transform. The full three-dimensional correlation function is then fitted directly with an analytic Lévy parameterization, allowing the source parameters $\alpha$, $\lambda$, $R_{\rm out}$, $R_{\rm side}$, and $R_{\rm long}$ to be extracted without reducing the source distribution to one-dimensional projections.

The method was applied to EPOS4 simulations of $0$--$10\%$ Au+Au collisions over the RHIC Beam Energy Scan energy range. The direct $C(\mathbf{q})$ analysis reproduces the main $m_T$- and collision-energy-dependent trends obtained previously from the $D(\boldsymbol{\rho})$-based analysis. Quantitatively, the extracted parameters show some systematic differences between the two approaches. The Lévy index $\alpha$ obtained from the $C(\mathbf{q})$ fits is systematically lower by about 15\%, while $R_{\rm out}$ remains very close to the previous results, with a difference of less than 5\%. The values of $R_{\rm side}$ and $R_{\rm long}$ differ by about 15\% and 20\%, respectively, while the corresponding $m_T$ and collision-energy dependences remain consistent between the two methods. The correlation strength $\lambda$ obtained from the $C(\mathbf{q})$ fits is is generally about 7\% larger, while its $m_T$ and collision-energy dependences are also preserved. The difference in $\alpha$ becomes more pronounced at the two lowest collision energies, 7.7 and 9.2~GeV, whereas the other fitted parameters and the quality of the $C(\mathbf{q})$ fits show no corresponding anomalous behavior. This indicates that the enhanced low-energy difference is specific to the extraction of the Lévy shape parameter rather than a general discrepancy between the two analyses.

We further compared the EPOS4 results with recent UrQMD calculations obtained using the $D(\boldsymbol{\rho})$-based method. The Lévy index $\alpha$ shows good agreement between the two models over the common collision-energy range, whereas the femtoscopic radii exhibit different excitation functions. In particular, $R_{\rm out}$ increases more strongly with collision energy in UrQMD than in EPOS4, while the energy dependence of $R_{\rm side}$ is more similar in the two models. As a consequence, the derived observables $R_{\rm diff}^{2}=R_{\rm out}^{2}-R_{\rm side}^{2}$ and $R_{\rm out}/R_{\rm side}$ show opposite energy dependences: they decrease with collision energy in EPOS4 but increase in UrQMD. These differences demonstrate that the excitation functions of out-side observables are sensitive to the underlying space-time evolution of the collision system.

Using
$\Delta\tau_{\rm eff}
\simeq
\frac{\sqrt{R_{\rm out}^{2}-R_{\rm side}^{2}}}{\beta_T}$ 
we find that the effective emission-time scale has only a weak collision-energy dependence in EPOS4, whereas it increases approximately linearly with $\sqrt{s_{NN}}$ in UrQMD. At fixed collision energy, $\Delta\tau_{\rm eff}$ decreases with increasing $k_T$ in both models, consistent with the reduction of the homogeneity region probed by higher-$k_T$ particles. Since $R_{\rm out}^{2}-R_{\rm side}^{2}$ also contains contributions from spatial and space-time correlations, $\Delta\tau_{\rm eff}$ should be regarded as an effective emission-time scale rather than a direct measurement of the global emission duration.

Overall, the comparison demonstrates that substantial energy dependence of femtoscopic observables can arise from differences in the underlying collision dynamics and space-time evolution. EPOS4 and UrQMD therefore provide complementary reference calculations for characterizing the range of excitation functions that can emerge from different model descriptions. Such model comparisons are important when interpreting non-monotonic structures in experimental femtoscopic observables and provide a necessary baseline for future investigations of possible phase-transition and critical-point-related phenomena.

The present work establishes a three-dimensional, model-comparable framework for studying femtoscopic excitation functions. Future extensions to experimentally measured correlation functions and to model calculations with different equations of state and dynamical scenarios will enable a more systematic assessment of the sensitivity of Lévy source parameters and out-side observables to the underlying QCD dynamics. Such comparisons will be essential for establishing robust baselines and for assessing whether non-monotonic structures observed in experimental excitation functions can be attributed to generic dynamical effects or may be associated with the QCD phase diagram.

\section*{Acknowledgments}
This research was funded by the NKFIH grants TKP2021-NKTA-64, PD-146589, K-146913, K-138136, and NKKP ADVANCED 152097.

\section*{ORCID}

\noindent Yan Huang - \url{https://orcid.org/0000-0002-1499-6051}

\noindent Máté Csanád - \url{https://orcid.org/0000-0002-3154-6925}


\begin{thebibliography}{0}

\bibitem{Tumasyan2024LHC}
A.~Tumasyan {\it et~al.} (CMS Collaboration), {Two-particle Bose-Einstein correlations and their L\'evy parameters in PbPb collisions at $\sqrt{s_{NN}} = 5.02$ TeV}, Phys. Rev. C 109 (2024) 024914.
\newblock \href {https://doi.org/10.1103/PhysRevC.109.024914} {\path{doi:10.1103/PhysRevC.109.024914}}.

\bibitem{Kincses2024RHIC}
D.~Kincses, {Pion interferometry with L\'evy-stable sources in $\sqrt{s_{NN}} = 200$ GeV Au+Au collisions at STAR}, Universe 10 (2024) 102.
\newblock \href {https://doi.org/10.3390/universe10030102} {\path{doi:10.3390/universe10030102}}.


\bibitem{Porfy2024SPS}
B.~P\'orfy, {Femtoscopy analysis in small systems at NA61/SHINE}, EPJ Web Conf. 296 (2024) 06004.
\newblock \href {https://doi.org/10.1051/epjconf/202429606004} {\path{doi:10.1051/epjconf/202429606004}}.

\bibitem{Csorgo2006QCD}
T.~Cs\"org\H{o}, S.~Hegyi, T.~Novak, and W.~A.~Zajc,
{Bose-Einstein or HBT correlation signature of a second order QCD phase transition},
AIP Conf. Proc. 828 (2006) 525--532.
\newblock \href {https://doi.org/10.1063/1.2197465} {\path{doi:10.1063/1.2197465}}.


\bibitem{Kincses2026}
D.~Kincses, E.~\'Arp\'asi, L.~Kov\'acs, M.~Nagy, M.~Csan\'ad,
{Three-dimensional sizes and shapes of pion emission in heavy-ion collisions},
Eur.\ Phys.\ J.\ C \textbf{86} (2026) 333.
\newblock \href {https://link.springer.com/article/10.1140/epjc/s10052-026-15526-1} {\path{doi:10.1140/epjc/s10052-026-15526-1}}.


\bibitem{Csanád_2025}
M.~Csanád, D.~Kincses, {Investigating the excitation function of HBT radii for Lévy-stable sources}, Journal of Physics G: Nuclear and Particle Physics 52~(2)  025102.
\newblock \href {https://doi.org/10.1088/1361-6471/ada04f} {\path{doi:10.1088/1361-6471/ada04f}}.

\bibitem{Huang2026}
Y.~Huang, M.~Moln\'ar, D.~Kincses, M.~Csan\'ad,
{Excitation function of femtoscopic L\'evy source parameters of pion pairs in EPOS4},
Phys.\ Lett.\ B \textbf{876} (2026) 140423.
\newblock \href {https://doi.org/10.1016/j.physletb.2026.140423} {\path{doi:10.1016/j.physletb.2026.140423}}.



\bibitem{Kincses:2024lnv}
D.~Kincses, M.~Nagy, M.~Csan{\'a}d, {L{\'e}vy walk of pions in heavy-ion collisions}, Commun. Phys. 8~(1) (2025) 55.
\newblock \href {https://doi.org/10.1038/s42005-025-01973-x} {\path{doi:10.1038/s42005-025-01973-x}}.


\bibitem{Nolan2013LevyEq}
J.~P.~Nolan,
{Multivariate elliptically contoured stable distributions: theory and estimation},
Comput.\ Stat.\ \textbf{28} (2013) 2067--2089.
\newblock \href {https://link.springer.com/article/10.1007/s00180-013-0396-7} {\path{doi:10.1007/s00180-013-0396-7}}.



\bibitem{fftw}
Fastest Fourier Transform software: 
\newblock \href {https://www.fftw.org/} {\path{https://www.fftw.org/}}.



\bibitem{Molnar2026}
M.~Moln\'ar, D.~Kincses, M.~Csan\'ad,
{Pion emission source shape in UrQMD Au+Au collisions at STAR energies},
arXiv:2605.15717 [nucl-th] (2026).
\newblock \href {https://doi.org/10.48550/arXiv.2605.15717} {\path{doi:10.48550/arXiv.2605.15717}}.





\end{thebibliography}
\end{document}